\documentclass[journal,twoside,web]{ieeecolor}
\usepackage{tmi}
\usepackage{cite}
\usepackage{amsmath,amssymb,amsfonts}
\usepackage{algorithmic}
\usepackage{graphicx}
\usepackage{booktabs}
\usepackage{multirow}
\usepackage{textcomp}
\usepackage{hyperref}
\hypersetup{hypertex=true,
            colorlinks=true,
            linkcolor=blue,
            anchorcolor=blue,
            citecolor=blue}
\def\BibTeX{{\rm B\kern-.05em{\sc i\kern-.025em b}\kern-.08em
    T\kern-.1667em\lower.7ex\hbox{E}\kern-.125emX}}
\begin{document}
\title{FANCL: Feature-Guided Attention Network with Curriculum Learning for Brain Metastases Segmentation}
\author{Zijiang Liu, Xiaoyu Liu, Linhao Qu, Yonghong Shi
\thanks{This work was supported by the National Natural Science Foundation of China under Grant 82072021. (Corresponding author: Yonghong Shi)}
\thanks{All the authors are with Digital Medical Research Center, School of Basic Medical Science, Fudan University, Shanghai 200032, China. They are also with Shanghai Key Lab of Medical Image Computing and Computer Assisted Intervention, Shanghai 200032, China. (email: liuzj23@m.fudan.edu.cn, liuxiaoyu21@m.fudan.edu.cn, lhqu20@fudan.edu.cn, yonghongshi@fudan.edu.cn).}}

\maketitle

\begin{abstract}
Accurate segmentation of brain metastases (BMs) in MR image is crucial for the diagnosis and follow-up of patients. Methods based on deep convolutional neural networks (CNNs) have achieved high segmentation performance. However, due to the loss of critical feature information caused by convolutional and pooling operations, CNNs still face great challenges in small BMs segmentation. Besides, BMs are irregular and easily confused with healthy tissues, which makes it difficult for the model to effectively learn tumor structure during training. To address these issues, this paper proposes a novel model called feature-guided attention network with curriculum learning (FANCL). Based on CNNs, FANCL utilizes the input image and its feature to establish the intrinsic connections between metastases of different sizes, which can effectively compensate for the loss of high-level feature from small tumors with the information of large tumors. Furthermore, FANCL applies the voxel-level curriculum learning strategy to help the model gradually learn the structure and details of BMs. And baseline models of varying depths are employed as curriculum-mining networks for organizing the curriculum progression. The evaluation results on the BraTS-METS 2023 dataset indicate that FANCL significantly improves the segmentation performance, confirming the effectiveness of our method.
\end{abstract}


\section{INTRODUCTION}
\label{sec:introduction}
\IEEEPARstart{I}{n} the central nervous system, brain metastases (BMs) are the most common malignant tumors \cite{bmsbg0}. Patients with BMs require accurate diagnosis and effective treatment; otherwise, their prognosis is poor \cite{bmsbg1}. Currently, one of the main treatments for BMs is radiation therapy, which necessitates precise detection and segmentation of BMs \cite{bmsbg2}. And BMs are primarily manually segmented by human experts in MR images, which is both time-consuming and difficult to reproduce \cite{bmsbg0}, \cite{bmsbg3}, \cite{bmsbg4}. Therefore, there is an urgent need for automatic methods to address the challenges of BMs segmentation.

\begin{figure}[t]
    \centering
    \includegraphics[width=1\linewidth]{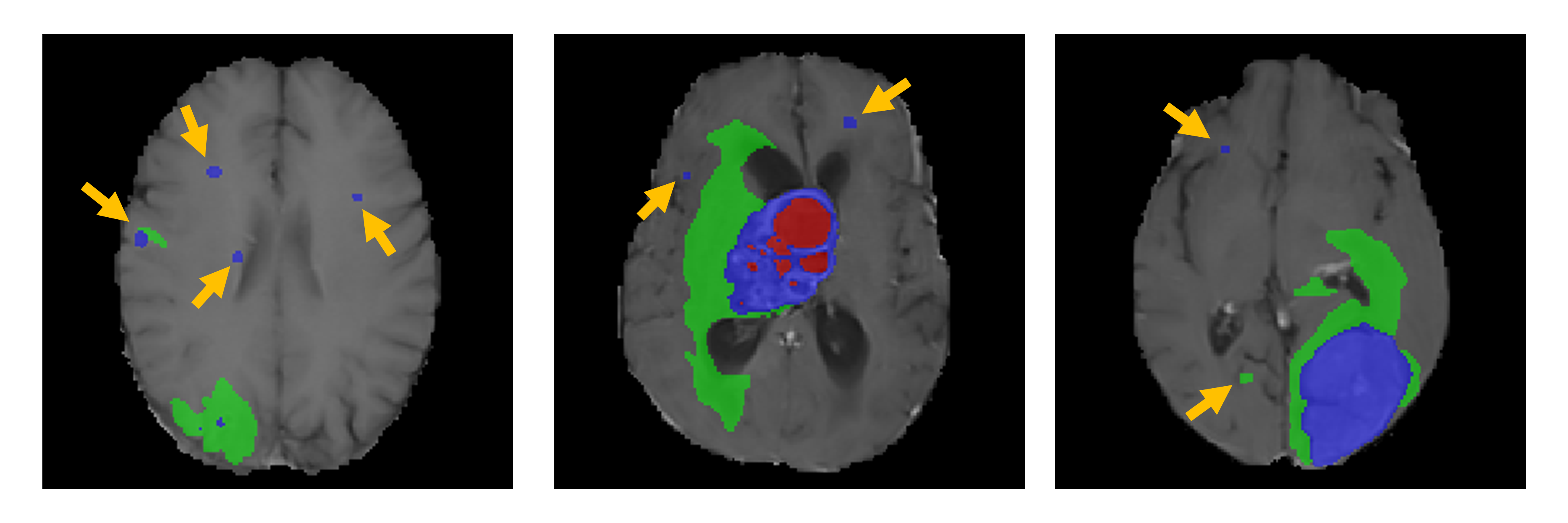}
    \caption{Illustration of visual examples of BMs. Here, red, green, and blue represent the nonenhancing tumor, peritumoral edema and enhancing tumor, respectively. Small tumors are indicated by yellow arrows. The high heterogeneity in size, shape and density of metastases, along with the complex anatomical structure of the brain, makes the segmentation task extremely difficult.}
    \label{vis}
\end{figure}

\begin{figure}[htb]
    \centering
    \includegraphics[width=1\linewidth]{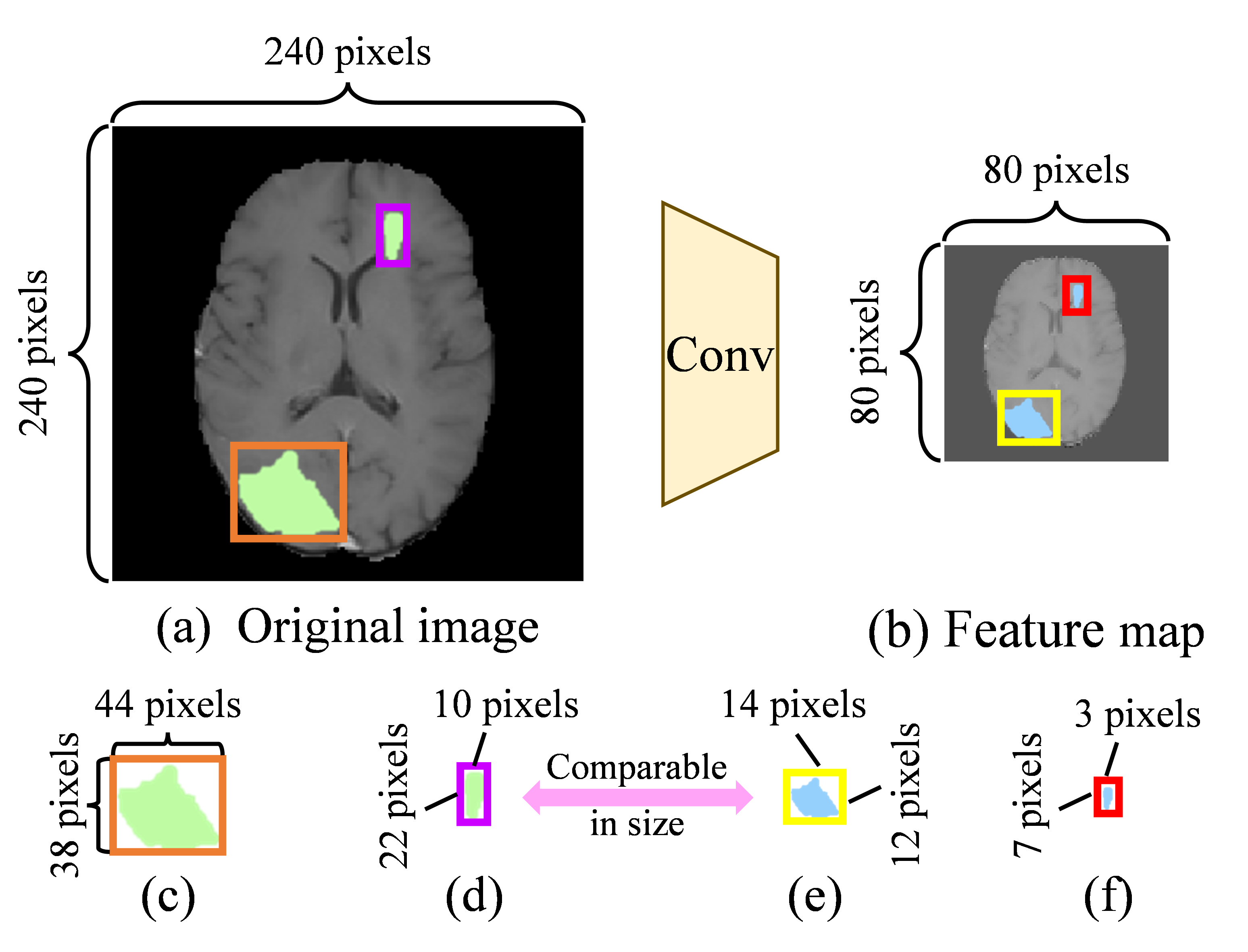}
    \caption{Visualization of the original image and its corresponding convolutional feature map. Here, \textbf{Conv} is a $3\times 3\times 3$ convolutional layer with the stride of 2. \textbf{(a)} $240\times 240$ 2D slice of the original image. \textbf{(b)} $80\times 80$ feature map slice of the original image generated by \textbf{Conv}. \textbf{(c)} The larger tumor in the original image. \textbf{(d)} The smaller tumor in the original image. \textbf{(e)} The feature map of the larger tumor. \textbf{(f)} The feature map of the smaller tumor. It is evident that there is a significant difference in tumor size between (c) and (d), but a small difference between (d) and (e). This indicates that the large tumors in the feature maps and small tumors in the original images have more comparable sizes. Meanwhile, (f) shows that the feature map of small tumors obtained through convolutional operation is very small, making it difficult to provide sufficient information.}
    \label{FGA_exp}
\end{figure}

In recent years, automatic methods based on convolutional neural networks (CNNs) have demonstrated good performance in medical image segmentation \cite{vnet}, \cite{cnn1}, \cite{cnn2}, \cite{cnn3}. Specifically, many excellent models are built on the encoder-decoder architecture \cite{cnn1}, \cite{cnn3}, among which U-Net \cite{unet} is one of the most typical and widely used network. On this basis, Fabian Isensee et al. proposed nnU-Net \cite{nnunet1}, which fully leverages the advantages of deep CNNs and achieves outstanding performance in various segmentation domains \cite{nnunet2}, \cite{nnunet3}. However, it is worth noting that BMs in MR images exhibit the following characteristics \cite{bmsbg2}, \cite{bmsct1}, \cite{bmsct2}: $\textcircled{1}$ BMs vary in size, and there are many small lesions. $\textcircled{2}$ The number and distribution of BMs vary among different patients. $\textcircled{3}$ BMs have complex and irregular structures. Fig. \ref{vis} illustrates some visual examples. Typically, there are numerous small tumors (indicated by yellow arrows) and large tumors with complex structures in MR images of patients with BMs, and their feature information is likely to be lost during the convolution process. Meanwhile, due to the inherent locality of convolutional operation, it is difficult for CNNs to establish long-range dependencies. Thus, many researchers consider using attention mechanism to tackle these problems \cite{attcnn1}, \cite{attcnn2}, \cite{attcnn3}, \cite{attincnn}. For example, Swin UNETR \cite{swinunetr} combines the U-shaped network architecture with Swin Transformer to improve the performance of brain glioma segmentation. Yang et al. \cite{bratsTU} adapt a hybrid model of CNN-Transformer called TransUNet for the segmentation of BMs to improve the accuracy of small tumors and achieve good results, which demonstrates the effectiveness of the attention mechanism in the segmentation of small metastases.

However, the inherent characteristics of BMs still pose significant challenges for precise segmentation. Some noise and small blood vessels appear very similar to BMs in MR images \cite{bmsbg2}. Meanwhile, infiltrative growth and partial volume effects further obscure the boundaries of lesions, particularly for small metastases \cite{bmsbg2}. These combined factors result in even advanced segmentation techniques struggling to achieve perfect segmentation of BMs, highlighting the ongoing need for innovation in this field. Recently, curriculum learning (CL) \cite{cl} has achieved many significant results in the field of medical image segmentation \cite{clspirit}, \cite{clinmi1}, \cite{clinmi2}. Their commonality lies in the use of CL to tackle complex segmentation tasks and further enhance segmentation accuracy. CL strategy suggests that the training steps of the model proceed in a meaningful order from easy to difficult, making the model have better performance.

Inspired by the above analysis and references \cite{afma} and \cite{gre}, we propose a novel model called Feature-guided Attention Network with Curriculum Learning (FANCL). Firstly, noticing that compared to normal tissues, tumors of different sizes often exhibit similar imaging characteristics (such as contrast and texture). This suggests that we should consider using the intrinsic correlation between large and small BMs to compensate for the information loss of small tumors in feature propagation. Meanwhile, we focus on the original image and feature space, discovering that the size differences between large tumors in small-sized feature maps and small tumors in original images are relatively small (see Fig. \ref{FGA_exp}), which makes it possible to find intrinsic correlation. In view of this, we present feature-guided attention mechanism that obtains the correlation between BMs of different sizes by computing the cross-correlation matrix (denoted as FGA) between the input image and the intermediate output feature. Then the similarity can be used to correct the segmentation results of small tumors. Moreover, in order to improve the accuracy of the generated FGA, we also calculate the golden FGA (denoted as GFGA) based on the ground truth mask to supervise the acquisition of FGA. Secondly, in order to enable the model to fully learn the tumor structure and details, this paper employs the CL strategy, reorganizing the input data in a meaningful order to provide progressive curriculum guidance for training. Furthermore, considering that BMs segmentation is the voxel-level prediction task, easy and difficult voxels exist simultaneously in each sample. Hence, sample-level curriculum cannot provide sufficient guidance for model training. In view of this, we propose a voxel-level CL strategy to improve the performance of our model. Additionally, combining the advantages of the methods introduced in \cite{clspirit} and \cite{gre}, we apply baseline models (nnU-Net) with different network depths as curriculum-mining networks to obtain curricula at different stages.

\begin{figure}[t]
    \centering
    \includegraphics[width=0.95\linewidth]{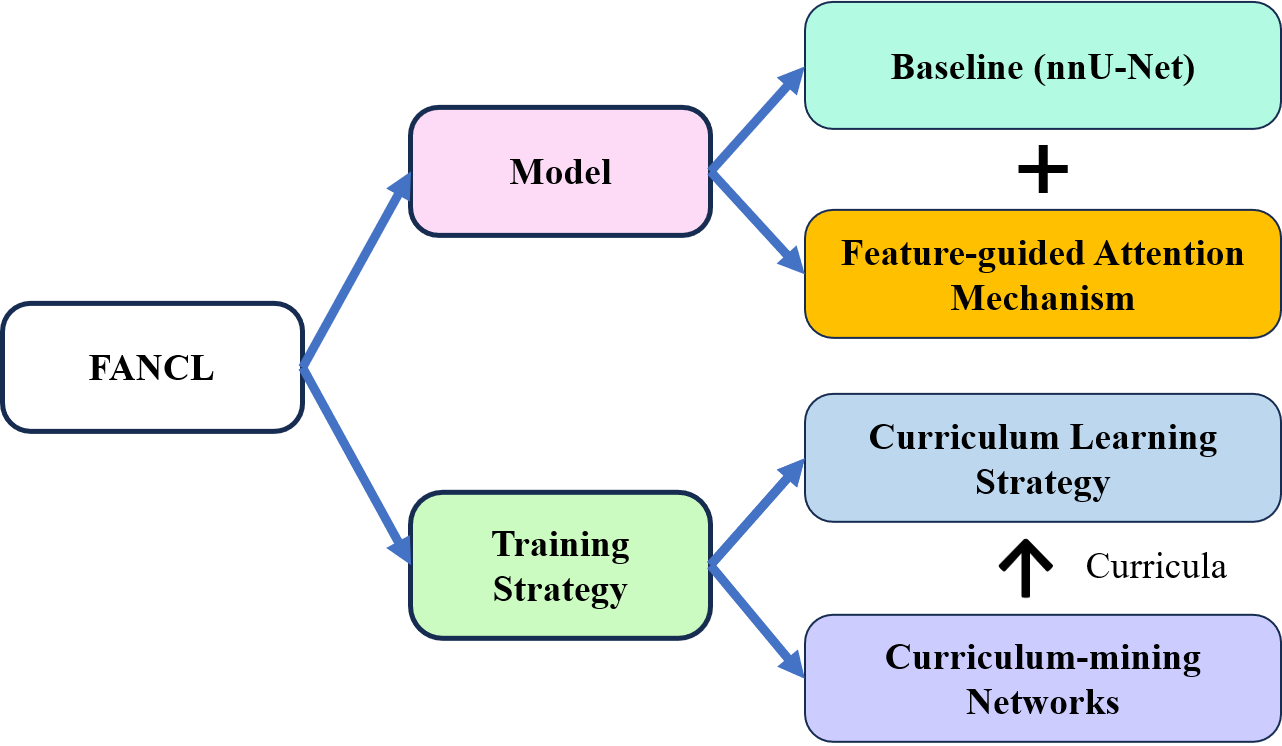}
    \caption{Architecture diagram of FANCL.}
    \label{FANCL}
\end{figure}

\begin{figure*}[t]
    \centering
    \includegraphics[width=1\textwidth]{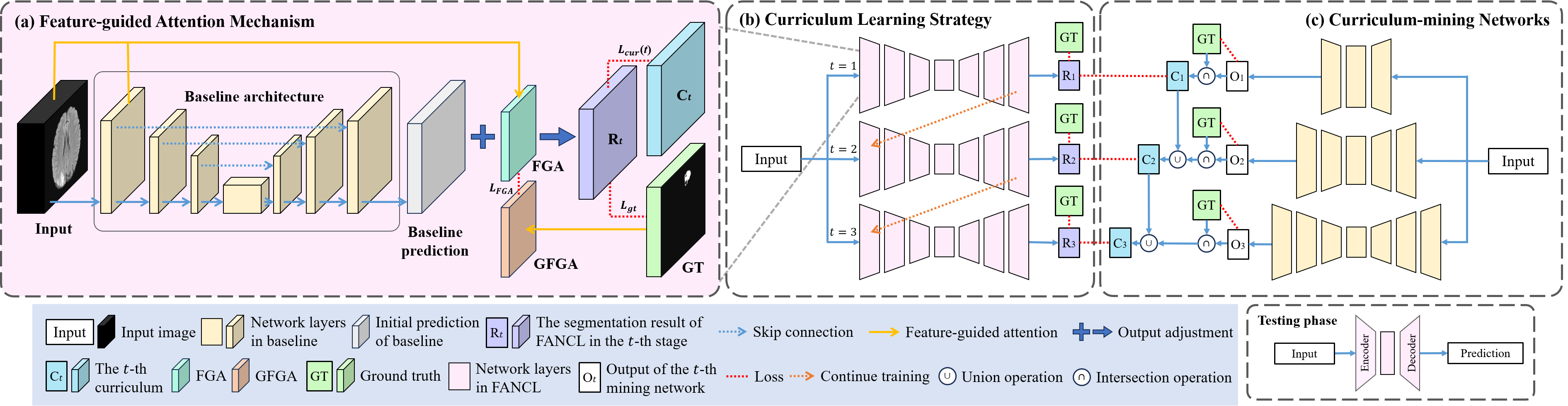}
    \caption{The framework of our methods. \textbf{(a)} The overview of our model, where the baseline architecture is nnU-Net. The feature-guided attention mechanism obtains FGA from the input image and feature map of baseline, and adjusts the initial prediction via FGA to obtain the modulated result $\mathrm{R}_t$. We also calculate the golden FGA (GFGA) through the ground truth (GT) for supervision. \textbf{(b)} Model training pipeline under the CL strategy. In the $t$-th stage, model is supervised by $\mathrm{C}_t$ obtained in (c) and GT simultaneously. When $t\geq 2$, the model continues training from the previous stage. \textbf{(c)} Curriculum-mining strategy, which extracts different feature information in the input image through networks of different depths, and then forms curricula $\mathrm{C}_t$ for different stages. Our model in (a) and the curriculum-mining networks in (c) are trained independently. Besides, in testing phase, the prediction can be obtained by simply inputting the image into our model.}
    \label{allinone}
\end{figure*}

The architecture diagram of FANCL is shown in the Fig. \ref{FANCL}, which consists of the three modules mentioned above. Furthermore, in order to demonstrate the effectiveness of FANCL, we evaluate it on the publicly available dataset BraTS-METS 2023, and the results show that FANCL significantly improves the segmentation accuracy. The main contributions in this work are presented as follows:
\begin{enumerate}
    \item We successfully apply the feature-guided attention mechanism to the segmentation task of BMs. This method utilizes feature map to compensate for the information loss of small tumor in convolutional operations.
    \item To the best of our knowledge, it is the first time to use voxel-level CL strategy for 3D medical image segmentation, which improves the segmentation accuracy.
    \item In order to design the curriculum progress that meets the model performance, we use baseline models with different depths as curriculum-mining networks. This strategy effectively avoids introducing complex network structures.
    \item We compare the proposed model with existing methods on the BraTS-METS 2023 dataset and demonstrate the effectiveness of FANCL through evaluation metrics and visualization results.
\end{enumerate}


\section{RELATED WORK}
\subsection{Medical Image Segmentation Based on Convolutional Neural Networks}
With the rapid development of CNNs in recent years, many researchers have introduced convolutional networks into the field of medical image segmentation, resulting in numerous successful models \cite{cnn2}, \cite{cnn3}, \cite{lxy}, \cite{3duxnet}. Therein, Fully Convolutional Networks (FCNs) have gained significant attention due to the ability to flexibly handle input images of variable sizes. In particular, U-Net \cite{unet}, a well-known and widely used model, is an improvement based on FCNs, which incorporates skip connection and a symmetrical U-shaped encoder-decoder structure, and achieves outstanding performance. Subsequently, various variants have been developed for different challenges, such as V-Net \cite{vnet}, nnU-Net \cite{nnunet1}, 3D U-Net \cite{3dunet} and UNet++ \cite{unetpp}, all of which yield good results. Among them, nnU-Net brilliantly completes multiple medical image segmentation tasks and is widely used in brain tumor segmentation \cite{nnunet2}, \cite{nnunetinbts}. Therefore, this paper utilizes nnU-Net as the baseline model to deal with the segmentation problem of BMs.

\subsection{Attention Mechanism}
In order to make the network focus on the most important part of the input data, the attention mechanism is integrated into the segmentation model. This mechanism takes all elements in the sequence into account during calculation, which gives the model a global perspective and a more comprehensive understanding of the lesion structure. As is well known, CNNs excel at extracting local feature but struggle to capture global information, and thus the introduction of attention mechanism can alleviate this issue. There have been many attempts in this regard \cite{swinunetr}, \cite{unetpp}, \cite{unetr}, \cite{attentionunet}. For example, Jieneng Chen et al. propose TransUNet \cite{transunet}, skillfully combining the globality of Transformer with the locality of U-Net. Furthermore, based on the U-shaped network architecture, Swin UNETR \cite{swinunetr} applies Swin Transformer as the complete encoder, and delivers features of different resolutions to an FCN-based decoder via skip connection. Swin UNETR achieves better performance due to its special model design. The study in \cite{attincnn} utilizes position and channel-wise attention modules to generate attention feature maps at different scales, which help filter noise out and highlight the target area to improve the accuracy of medical image segmentation. In this paper, we propose feature-guided attention mechanism that effectively leverages the feature information of large tumors to guide the segmentation of small ones, thereby enhancing the model performance.

\subsection{Curriculum Learning Strategy}
Inspired by cognitive theory, CL strategy is first proposed by Bengio et al. in \cite{cl}. CL strategy points out that training the model in a manner that gradually increases the difficulty of the input data will outperform the common training method based on random sequential data. At present, some researchers have successfully applied the CL strategy to the field of medical images. In \cite{clinmi1}, a three-stage CL method is used to segment liver tumor. Peng Tang et al. \cite{clspirit} adopt CL approach with step-wise training strategy in skin lesion segmentation to address the issue of overfitting caused by sample imbalance. Although these studies have achieved promising results, they neglect the fact that segmentation is a voxel-level prediction work, which means that the training data provided to the model contains both easy-to-segment and difficult-to-segment voxels. In view of this, we apply a voxel-level CL strategy in this work to tackle the challenge of BMs segmentation.

\section{METHOD}
The overview of our model is shown in Fig. \ref{allinone}(a). The feature-guided attention mechanism takes the original image and the feature map from shallow encoding layer as input, and its output (FGA) is combined with the initial prediction of baseline as the final segmentation result ($\mathrm{R}_t$). Specifically, our approach utilizes the input to calculate the cross-correlation matrix (i.e., FGA) to quantify the intrinsic relationship between tumors of different sizes (see Section 3.1), and then applies the obtained FGA to adjust the output from the decoder (Section 3.2). Meanwhile, we also compute GFGA via the ground truth mask to guide the generation of FGA (Section 3.3).

Besides, the training pipeline of our model under the voxel-level CL strategy is shown in Fig. \ref{allinone}(b), where the curricula of each stage are provided by the curriculum-mining networks. The content related to CL strategy and the mining network is discussed in section 3.4. Finally, Section 3.5 introduces the loss function used in the model training process.
\subsection{Feature-Guided Attention}
\begin{figure}[htbp]
    \centering
    \includegraphics[width=1\linewidth]{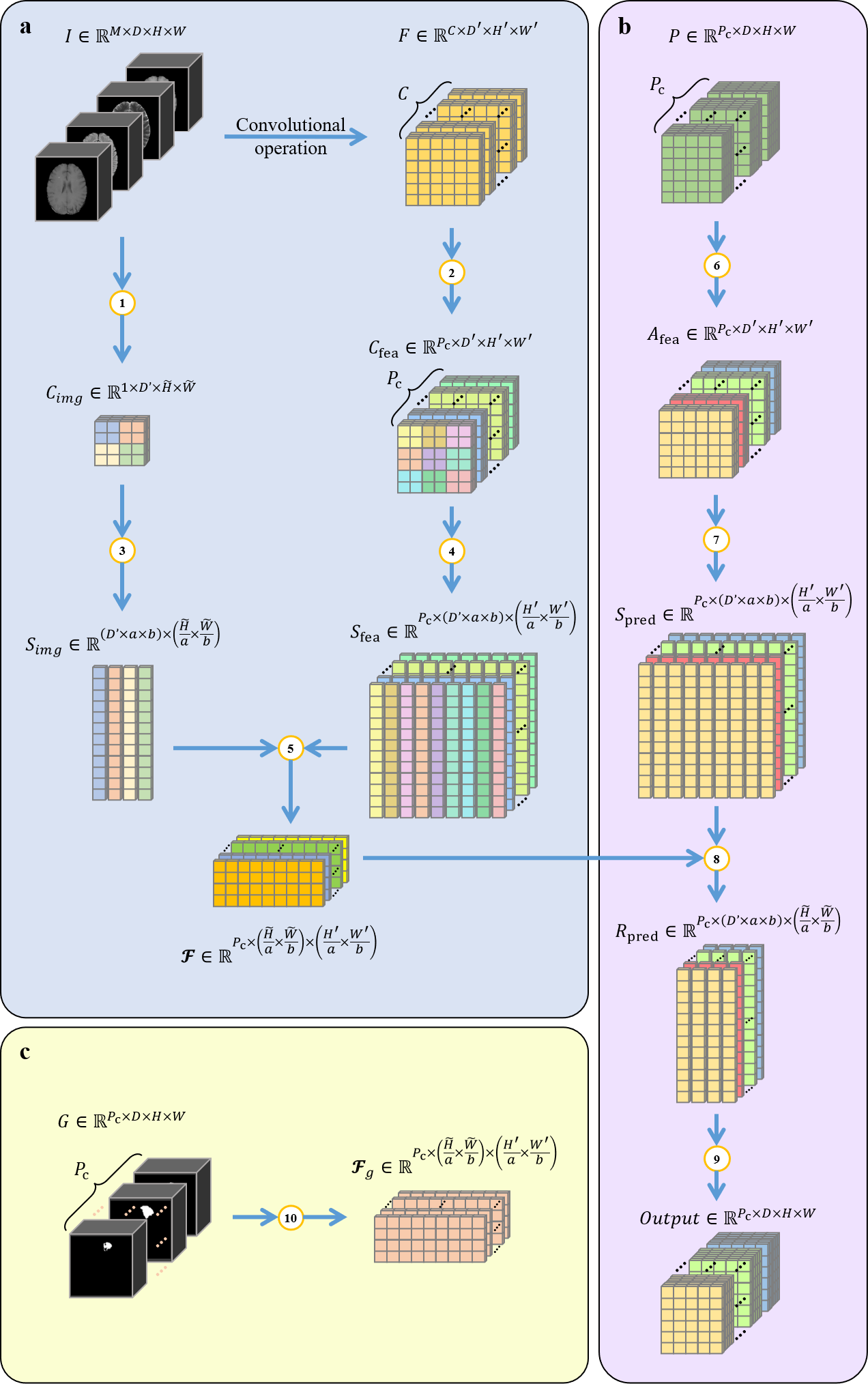}
    \caption{Feature-guided attention mechanism. \textbf{a}: The calculation process of FGA. \textbf{b}: Use the FGA obtained in \textbf{a} to adjust the initial prediction of baseline. \textbf{c}: The generation process of GFGA.}
    \label{FGA_steps}
\end{figure}

In this section, we use the original image and feature map to quantify the relationship between BMs of different sizes. Fig. \ref{FGA_steps}(a) shows the specific process. The input image is denoted as $I\in \mathbb{R}^{M\times D\times H\times W}$, where $M$ represents the number of modalities in the input MR image, and $D, H, W$ represent the depth, height and width of the image, respectively. The shallow feature map output by baseline encoder is $F\in \mathbb{R}^{C\times D’\times H’\times W’}$. Therein, $C$ represents the number of channels, and the size of the feature is determined by the convolutional network. Our model uses the original image $I$ and the first-layer feature map $F$ as input to calculate FGA. The detailed process is described as follows.

\textbf{Step~1}: The convolutional operation $\mathbf{Conv_{img}}$ is used to preprocess the input image and compress the second dimension to the same size as $F$, i.e. $D’$. This process is recorded as
\begin{equation}
    C_{img}= \mathbf{Conv_{img}}(I),
\end{equation}
where $C_{img}\in\mathbb{R}^{1\times D’\times \widetilde{H}\times \widetilde{W}}$, $\widetilde{H}$ and $\widetilde{W}$ are the sizes after the convolutional operation. Specifically, $\mathbf{Conv_{img}}$ consists of a $7\times 7\times 7$ convolutional layer with 64 filters and a $3\times 3\times 3$ convolutional layer with 1 filter.

\textbf{Step~2}: The convolutional operation $\mathbf{Conv_{fea}}$ is a $1\times 1\times 1$ convolutional layer with $P_c$ filters, where $P_c$ represents the number of predicted categories. $\mathbf{Conv_{fea}}$ converts the feature map $F$ into each category features:
\begin{equation}
    C_{fea}=\mathbf{Conv_{fea}}(F),
\end{equation}
where $C_{fea}\in\mathbb{R}^{P_c\times D’\times H’\times W’}$. In addition, we denote $C^k_{fea}\in\mathbb{R}^{1\times D’\times H’\times W’}, k=\{1,2,\cdots,P_c\}$ as the feature containing the $k$-th predicted category information.

\textbf{Step~3}: Next, $C_{img}$ derived in \textbf{Step~1} is divided into a series of flattened patches of fixed size:
\begin{equation}
    S_{img}=\phi(C_{img},a,b).
\end{equation}
This step can reshape $C_{img}\in\mathbb{R}^{1\times D’\times \widetilde{H}\times \widetilde{W}}$ into a 2D map $S_{img}\in\mathbb{R}^{(D’\times a\times b)\times (\frac{\widetilde{H}}{a}\times \frac{\widetilde{W}}{b})}$ through the partition operation $\phi$, where each column vector $s\in\mathbb{R}^{(D’\times a\times b)\times 1}$ represents a patch of size $1\times  D’\times a\times b$ from $C_{img}$. Note that $a$ and $b$ should be divisible by $\widetilde{H}$ and $\widetilde{W}$ respectively. For this purpose, bottom-right padding can be applied on $C_{img}$ if needed. The details of partition operation can refer to \cite{partition}, and we make reasonable extensions on this basis.

\textbf{Step~4}: Similar to \textbf{Step~3}, we split each $ C^k_{fea}$ into a series of long strips:
\begin{equation}
    S^k_{fea}=\phi(C^k_{fea},a,b),
\end{equation}
where $S^k_{fea}\in\mathbb{R}^{(D’\times a\times b)\times (\frac{H’}{a}\times \frac{W’}{b})}, k=\{1,2,\cdots,P_c\}$. Then concatenate them together to obtain 
\begin{equation}
    S_{fea}=\mathbf{Concat}_{k=1}^{P_c}(S^k_{fea}).
\end{equation}
Therein, $\mathbf{Concat}$ represents the concatenation operation and $S_{fea}\in\mathbb{R}^{P_c\times (D’\times a\times b)\times (\frac{H’}{a}\times \frac{W’}{b})}$.

\textbf{Step~5}: Calculate the dot product of $S_{img}$ and each $S^k_{fea}$ to obtain the correlation between the original image and the feature map of each predicted category. The process is denoted as:
\begin{equation}
    \mathcal{F}^k=(S_{img})^{\mathrm{T}}\otimes S^k_{fea},
\end{equation}
\begin{equation}\label{FGA}
    \mathcal{F}=\mathbf{Concat}_{k=1}^{P_c}(\mathcal{F}^k),
\end{equation}
where $\mathcal{F}^k\in \mathbb{R}^{(\frac{\widetilde{H}}{a}\times \frac{\widetilde{W}}{b})\times (\frac{H’}{a}\times \frac{W’}{b})}$, $\mathcal{F}\in \mathbb{R}^{P_c\times (\frac{\widetilde{H}}{a}\times \frac{\widetilde{W}}{b})\times (\frac{H’}{a}\times \frac{W’}{b})}$, $\otimes$ represents the dot product and $(S_{img})^{\mathrm{T}}$  is the transposed matrix of  $S_{img}$.

\subsection{Output Adjustment}
Next, as shown in Fig. \ref{FGA_steps}(b), the obtained attention matrix $\mathcal{F}$ is combined with the prediction output $P\in \mathbb{R}^{P_c\times D\times H\times W}$ of baseline to obtain the enhanced segmentation result. Therein, $P^k\in \mathbb{R}^{1\times D\times H\times W}, k=\{1,2,\cdots,P_c\}$ contains the predicted value for each voxel belonging to the $k$-th category.

\textbf{Step~6}: Average pooling operation $\mathbf{AP}$ is used to reshape the prediction $P$ to the same size as $C_{fea}\in\mathbb{R}^{P_c\times D’\times H’\times W’}$:
\begin{equation}
    A_{fea}=\mathbf{AP} (P),
\end{equation}
where $A_{fea}\in\mathbb{R}^{P_c\times D’\times H’\times W’}$. Meanwhile, $A_{fea}^k\in\mathbb{R}^{1\times D’\times H’\times W’}, k=\{1,2,\cdots,P_c\}$ contains the information related to the $k$-th category.

\textbf{Step~7}: Similar to \textbf{Step~4}, we divide each $A^k_{fea}$ into a series of flat patches, and then concatenate them in channel-wise: 
\begin{equation}
    S^k_{pred}=\phi(A^k_{fea},a,b),
\end{equation}
\begin{equation}
    S_{pred}=\mathbf{Concat}_{k=1}^{P_c}(S^k_{pred}),
\end{equation}
where $S^k_{pred}\in\mathbb{R}^{(D’\times a\times b)\times (\frac{H’}{a}\times \frac{W’}{b})}$ and $S_{pred}\in\mathbb{R}^{P_c\times (D’\times a\times b)\times (\frac{H’}{a}\times \frac{W’}{b})}$.

\textbf{Step~8}: Then, we combine $S_{pred}$ and the attention matrix $\mathcal{F}$ to adjust the output of the model:
\begin{equation}
    R^k_{pred}=S^k_{pred}\otimes (\mathcal{F}^k)^{\mathrm{T}},
\end{equation}
\begin{equation}
    R_{pred}=\mathbf{Concat}_{k=1}^{P_c}(R^k_{pred}),
\end{equation}
where $ R^k_{pred}\in \mathbb{R}^{(D’\times a\times b)\times(\frac{\widetilde{H}}{a}\times \frac{\widetilde{W}}{b})}, R_{pred}\in \mathbb{R}^{P_c\times(D’\times a\times b)\times(\frac{\widetilde{H}}{a}\times \frac{\widetilde{W}}{b})}$. Specifically, $\mathcal{F}$ contains the relationship between BMs of different sizes in the original image and the feature map , and $S_{pred}$ stands for the initial segmentation prediction of the model. Therefore, $R_{pred}$  represents the guidance of large tumor information for small tumor segmentation.

\textbf{Step~9}: Similar to the reverse process of \textbf{Steps~1} and \textbf{3}, we reshape $R_{pred}\in \mathbb{R}^{P_c\times(D’\times a\times b)\times(\frac{\widetilde{H}}{a}\times \frac{\widetilde{W}}{b})}$ to the same size as $P\in \mathbb{R}^{P_c\times D\times H\times W}$ via the inverse of partition operation $\phi$ and transpose convolution $\mathbf{TConv}$:
\begin{equation}
    O^k _{pred}=\mathbf{TConv}(\phi^{-1} (R^k_{pred},a,b)),
\end{equation}
\begin{equation}
    O_{pred}=\mathbf{Concat}_{k=1}^{P_c}(O^k_{pred}),
\end{equation}
where $O^k_{pred}\in \mathbb{R}^{1\times D\times H\times W}$, and $O_{pred}\in \mathbb{R}^{P_c\times D\times H\times W}$ represents correction information guided by features. The final output of the model is
\begin{equation}
    Output = P + O_{pred}.
\end{equation}

\subsection{Golden FGA}
To improve the generation quality of $\mathcal{F}$, we compute GFGA $\mathcal{F}_{g}\in \mathbb{R}^{P_c\times (\frac{\widetilde{H}}{a}\times \frac{\widetilde{W}}{b})\times (\frac{H’}{a}\times \frac{W’}{b})}$ based on the ground truth mask, as shown in Fig. \ref{FGA_steps}(c).

\textbf{Step~10}: Given the ground truth mask $G\in \mathbb{R}^{P_c\times D\times H\times W}$, where $G^k\in \mathbb{R}^{1\times D \times H\times W}$ contains the ground truth value related to the $k$-th category. Similar to \textbf{Steps~1-6}:
\begin{equation}
    C_{gt}= \mathbf{Conv_{gt}}(G),
\end{equation}
\begin{equation}
    A_{gt}=\mathbf{AP}(G),
\end{equation}
\begin{equation}
    \mathcal{F}_{g}^k=(\phi(C_{gt}^k,a,b))^{\mathrm{T}}\otimes\phi(A^k_{gt},a,b),
\end{equation}
\begin{equation}\label{FGA_gt}
    \mathcal{F}_{g}=\mathbf{Concat}_{k=1}^{P_c}(\mathcal{F}_{g}^k),
\end{equation}
where $C_{gt}\in\mathbb{R}^{P_c\times D’\times \widetilde{H}\times \widetilde{W}}$ is obtained through a convolutional operation similar to  $\mathbf{Conv_{img}}$ in \textbf{Step~1}, $A_{gt}\in\mathbb{R}^{P_c\times D’\times H’\times W’}$, and $\mathcal{F}_{g}\in \mathbb{R}^{P_c\times (\frac{\widetilde{H}}{a}\times \frac{\widetilde{W}}{b})\times (\frac{H’}{a}\times \frac{W’}{b})}$ embodies the golden correlation between the original image and the feature map.

Besides, it is worth mentioning that in this paper, the input image $I$ after baseline pre-processing has the same size as the output feature $F$ of the first layer in baseline encoder, i.e. $D=D’$, $H=H’$ and $W=W’$, which greatly simplifies parameter selection and calculation process.

\subsection{Voxel-Level Curriculum Learning and Curriculum-Mining Network}
\begin{figure}[htb]
    \centering
    \includegraphics[width=1\linewidth]{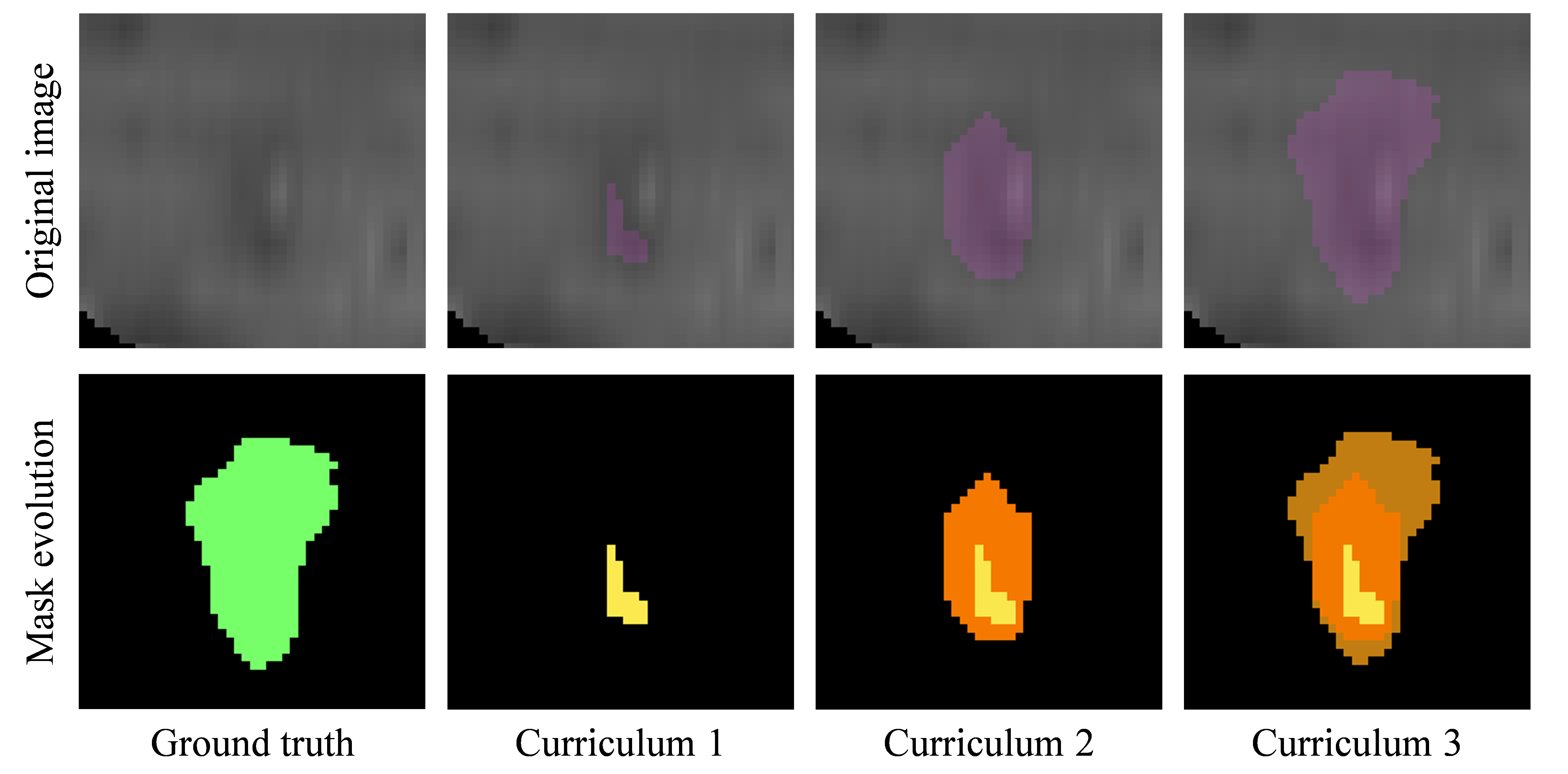}
    \caption{Illustration of the curriculum-mining process with an example. Green is the ground truth and pink is the mask corresponding to the curriculum. From yellow to orange to brown, the voxels contained in Curriculum 1 to 3 are increasingly difficult. The relationship between different curriculum stages is Curriculum 1 $\subset$ Curriculum 2 $\subset$ Curriculum 3.}
    \label{cur}
\end{figure}

In order to schedule the curriculum that conforms to the characteristics of BMs segmentation task, we assess the difficulty of the samples in voxel-wise. For each sample, the establishment of the curriculum is a process from an empty set to the entire ground truth mask. As shown in Fig. \ref{cur}, the curriculum region gradually expands with the increase of curriculum stages, and subsequent curricula always include previous curricula. During this process, the curriculum gradually transitions from easy-to-segment voxels to difficult-to-segment voxels, ultimately determining the distribution of curriculum at different stages. We can see that Curriculum 3 in Fig. \ref{cur} contains more difficult voxels than the other two curricula. Then, the progressive curricula are input into the model at specified time intervals to provide incremental guidance during training. In this work, we set up three curriculum stages to train our model. At the same time, considering that as the number of difficult voxels in an image increases, so does the difficulty of model learning, we set different learning epochs for each curriculum stage.

In addition, the difficulty of each voxel can be defined by visual information such as color and contrast, as well as deeper properties such as contextual relationships, which makes manual curriculum arrangement very challenging. Therefore, we utilize deep-learning networks to automatically mine curricula. In general, models of varying depths focus on different levels of information in the input data \cite{infowithd1}, \cite{infowithd2}, thereby obtaining different feature representations and output results. Shallow networks tend to extract low-level features to identify easy-to-segment areas, while deep networks prefer to use high-level features to mine more difficult parts. Moreover, it should be noted that designing a separate curriculum-mining network will bring additional labor and lead to model bias, which means that models of different structures may not consistently judge the difficulty of the same voxel. Therefore, this paper adopts a self-evaluation scheme, in which we apply baseline model as the backbone of curriculum-mining network, while setting up networks of different depths to obtain output results at various levels. As shown in Fig. \ref{allinone}(c), the outputs $\mathrm{O}_t$ of networks with different depths are used to organize curricula $\mathrm{C}_t$ at different stages, where $t=1,2,3$. The calculation method of $\mathrm{C}_t$ is as follows:
\begin{equation}
    \mathrm{C}_1 = \mathrm{O}_1 \cap \mathrm{GT},
\end{equation}
\begin{equation}
    \mathrm{C}_t = (\mathrm{O}_t \cap \mathrm{GT}) \cup \mathrm{C}_{t-1},~t\geq 2,
\end{equation}
where $\mathrm{GT}$ denotes the ground truth.

Next, we introduce the process of training the curriculum-mining network. In order to avoid interference from data with varying distributions, we employ the 5-fold cross-validation scheme to train the mining network on the entire dataset and collect prediction results of all data at different depths. This strategy helps obtain curriculum that align with the model performance and data characteristics.

\subsection{Loss Function}
The entire training process includes two aspects of loss, i.e., the segmentation loss, which evaluates the difference between the prediction result and the segmentation criteria, and the FGA loss, which calculates the gap between FGA and GFGA.

\subsubsection{Segmentation Loss}
The total segmentation loss function $L_{seg}$ consists of two parts: the loss $L_{cur}$ between the prediction and provided curriculum, and the loss $L_{gt}$ between the prediction and ground truth:
\begin{equation}
    L_{seg}=L_{cur}(t)+ L_{gt},
\end{equation}
where $t$ represents selecting the $t$-th curriculum for guidance. Since $L_{cur}(t)$ and $L_{gt}$ have the same form of expression, we adopt symbol $L$ to represent them below. Specifically, we use the same loss functions as baseline model, namely Dice loss and Cross Entropy loss:
\begin{equation}\label{loss}
    L=L_{Dice}+L_{CE},
\end{equation}
and
\begin{equation}
    L_{Dice}=-\frac{2}{C}\sum_{c=1}^{C}\frac{\sum_{(i,j,k)}y^c(i,j,k) \hat{y}^c(i,j,k)+\epsilon}{\sum_{(i,j,k)}\left(y^c(i,j,k)+\hat{y}^c(i,j,k)\right)+\epsilon },
\end{equation}
\begin{equation}
    L_{CE}=-\sum_{c=1}^{C}\sum_{(i,j,k)}y^c(i,j,k)\log\left(\hat{y}^c(i,j,k)\right),
\end{equation}
where $C$ stands for the number of categories, $c$ represents the $c$-th category, $(i,j,k)$ is the index of each voxel, $\hat{y}$ represents the prediction result of the model, and $y$ represents the segmentation criteria mask, which is the $t$-th curriculum for $L_{cur}(t)$  and the ground truth for $L_{gt}$.
In addition, due to applying baseline model as the backbone, the curriculum-mining network is also constrained by the loss function defined as (\ref{loss}), where $y$ represents the ground truth mask.

\subsubsection{FGA Loss}
In order to reduce the gap between FGA and GFGA, this study uses MSE loss:
\begin{equation}
    L_{FGA}=\frac{1}{CN}\sum_{c=1}^C\sum_{(i,j,k)}\left(\mathcal{F}^c(i,j,k)-\mathcal{F}_{g}^c(i,j,k)\right)^2,
\end{equation}
where $N$ represents the total number of voxels, and $\mathcal{F}$ and $\mathcal{F}_{g}$ are shown in equations (\ref{FGA}) and (\ref{FGA_gt}), respectively.

The entire loss function of FANCL is as follows:
\begin{equation}
    L_{FANCL}=L_{seg}+L_{FGA}.
\end{equation}

\section{Experiments}
\subsection{Dataset}
\begin{figure}[htbp]
    \centering
    \includegraphics[width=0.6\linewidth]{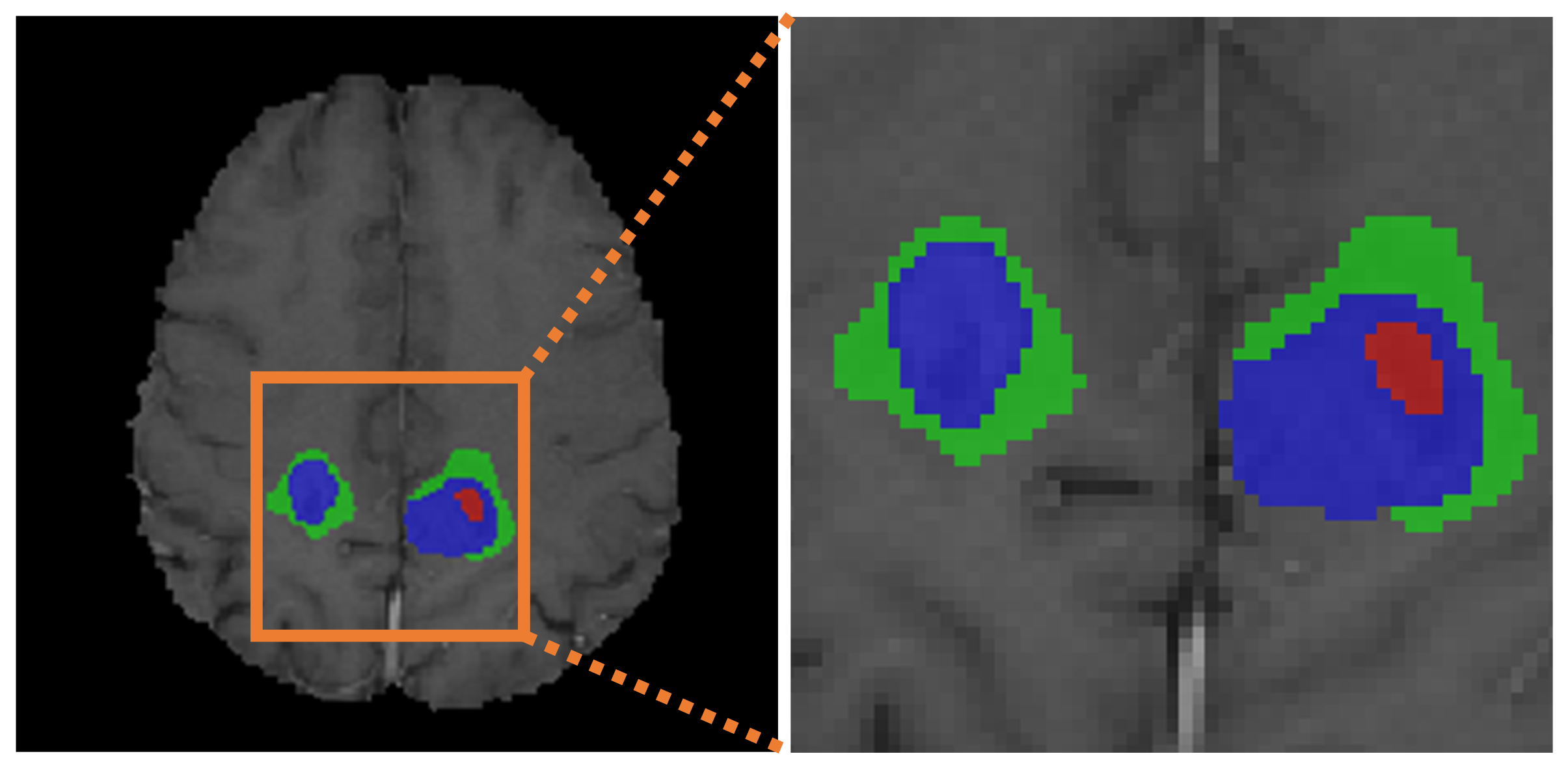}
    \caption{The segmentation labels of BMs. Red is the nonenhancing tumor (label 1), green is the peritumoral edema (label 2), and blue is the enhancing tumor (label 3).}
    \label{example}
\end{figure}

We evaluate our model using public dataset BraTS-METS 2023 \cite{brats}, which includes brain MR images of 4 modalities (T1, T1Gd, T2, T2-FLAIR) and corresponding BMs segmentation masks from 238 patients. The size of each image is (240, 240, 155), and segmentation label types are shown in Fig. \ref{example}: red is the nonenhancing tumor (label 1), green is the peritumoral edema (label 2), blue is the enhancing tumor (label 3), and background. Our goal is to segment the following three regions of interest: Enhancing Tumor (ET, label 3), Tumor Core (TC, label 1 + label 3), and Whole Tumor (WT, label 1 + label 2 + label 3).

\subsection{Evaluation Metric}
In this paper, we use Dice similarity coefficient (DSC) and 95\% Hausdorff distance (HD95) to evaluate the prediction results. The DSC value is calculated as follows:
$$\mathrm{DSC} = \frac{2\times TP}{2\times TP +FP +FN},$$
where $TP$ (True Positive) represents voxels accurately classified into their corresponding tumor categories, $FP$ (False Positive) denotes healthy tissue voxels erroneously identified as tumor regions, and $FN$ (False Negative) indicates tumor voxels mistakenly classified as healthy tissue. HD95 measures the 95-th percentile of the distances between the boundaries of predicted region and ground truth, excluding the unreasonable distances caused by some outliers. Its computing formula is as follows:
$$\mathrm{HD95} = \max\left\{\max_{x\in X}\min_{y\in Y}d(x,y), \max_{y\in Y}\min_{x\in X}d(x,y)\right\}_{\mathrm{95}},$$
where $d(x,y)$ stands for the Euclidean distance between x and y, $X$ and $Y$ represent the boundary point sets of the predicted mask and the ground truth, respectively. When both the prediction and the ground truth are empty, then $\mathrm{HD95}=0$; when only one is empty, the maximum $\mathrm{HD95}=373.1287$ is obtained; only when neither is empty, we use the above formula for calculation.

\subsection{Implementation Details}
In this work, 5-fold cross-validation scheme is applied to train the models. Specifically, we train the curriculum-mining networks for 100 epochs, and the rest of the networks for 500 epochs. In the CL strategy, we set up a total of 3 curriculum stages. In order to enable the model to fully learn difficult samples, the three stages of the curriculum are trained for 80, 80 and 340 epochs respectively. We use SGD optimizer with Nesterov momentum of 0.99 and poly learning rate schedule, in which the learning rate follows the polynomial decay with an exponent of 0.9. In addition, pre-processing and post-processing steps are consistent with those in nnU-Net. All experiments in this paper are performed by PyTorch on a single GPU NVIDIA A100.

\subsection{Results}
\subsubsection{Comparison to Baseline}
We use nnU-Net\footnote{\url{https://github.com/MIC-DKFZ/nnUNet}} as the baseline of our model. As shown in Table \ref{baseline_t}, FANCL improves DSC by 1.14\%, 1.39\% and 2.73\% compared with the baseline in the regions of ET, TC and WT respectively, and reduce HD95 by 3.07, 3.06 and 9.63 respectively, demonstrating the effectiveness of the proposed method.

\begin{table}[htbp]
    \centering
    \caption{Quantitative performance evaluation of baseline and our model using the metrics of DSC and HD95.}
    \label{baseline_t}
		\setlength{\tabcolsep}{1mm}{
    \begin{tabular}{c|cccc|cccc}
          \toprule
          \multirow{2}{*}{Model}& \multicolumn{4}{c|}{DSC~$\uparrow$} & \multicolumn{4}{c}{HD95~$\downarrow$} \\
          \cmidrule{2-9}
          & ET & TC & WT & Mean & ET & TC & WT & Mean \\
          \cmidrule{1-9}
          Baseline & 0.7449 & 0.8023 & 0.7754 & 0.7742 & 17.96 & 18.50 & 23.54 & 20.00 \\ 
          FANCL & \textbf{0.7563} & \textbf{0.8162} & \textbf{0.8027} & \textbf{0.7917} & \textbf{14.89} & \textbf{15.44} & \textbf{13.91} & \textbf{14.75} \\ 
          \bottomrule
    \end{tabular}}
\end{table}

\begin{figure}[htbp]
    \centering
    \includegraphics[width=1\linewidth]{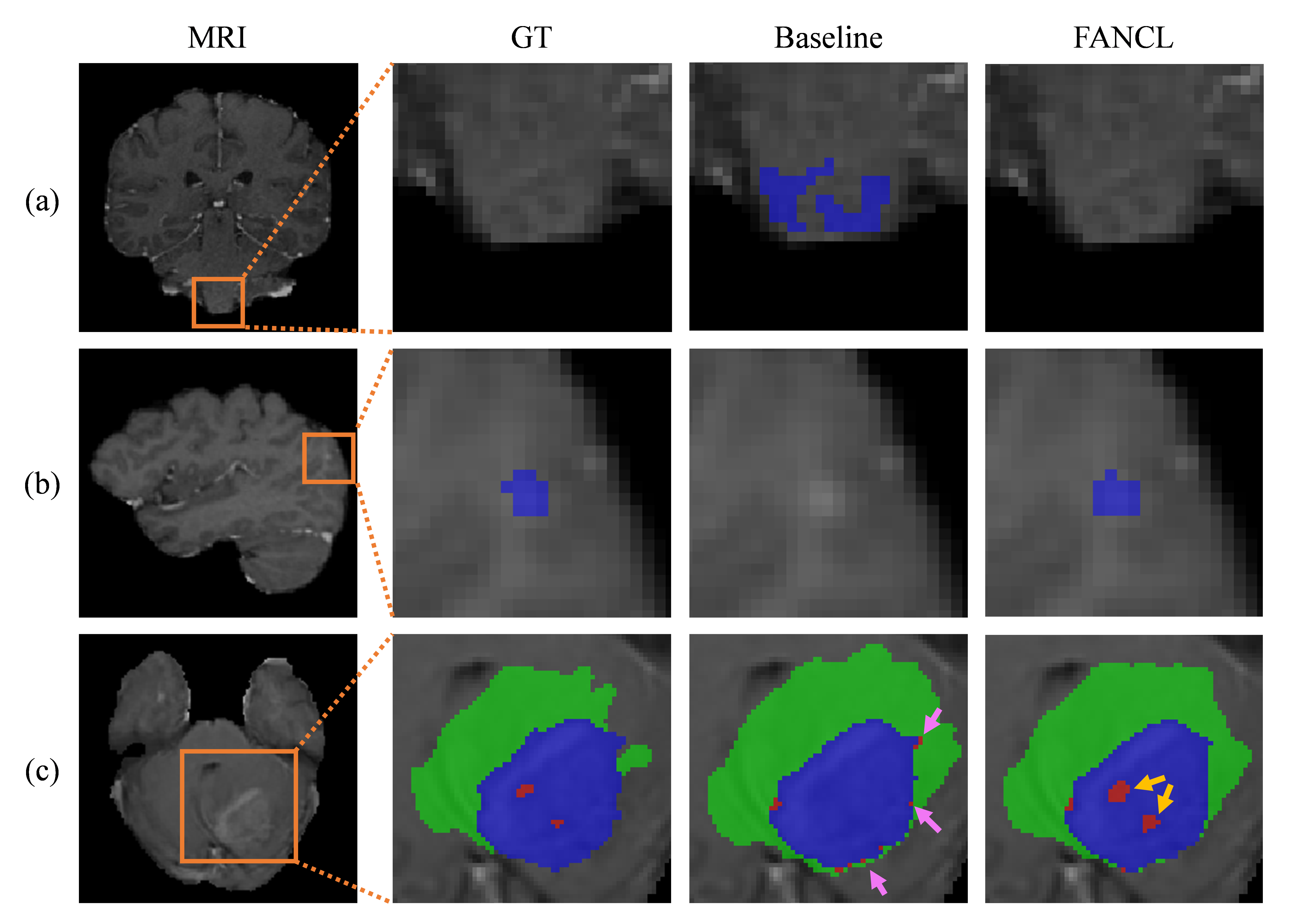}
    \caption{Comparative visualization of segmentation results between baseline and FANCL. First column: MR images (T1). Second column: the ground truth masks. Third column: baseline segmentation results. Fourth column: predictions of our proposed model.}
    \label{baseline_f}
\end{figure}

In Fig. \ref{baseline_f}, we show the segmentation results of baseline and FANCL. It can be observed that FANCL can effectively identify tumor regions, avoiding false positive predictions in baseline outputs (see (a)) and segmenting small tumors that are easily overlooked (see (b)). Besides, when facing complex BMs regions as shown in (c), FANCL optimizes the baseline prediction of the tumor structure (the shape of peritumoral edema is more consistent with that in the ground truth and mislabeled nonenhancing tumors are reduced, as indicated by pink arrows), and accurately predicts the small nonenhancing tumors in the center of the tumor region (yellow arrows).

\subsubsection{Comparison to Existing Models}
In this section, we compare FANCL with three excellent methods in the field of medical image segmentation: 3D U-Net \cite{3dunet}, 3D UX-Net \cite{3duxnet} and Swin UNETR\cite{swinunetr}, with the results presented in Table \ref{exist_t}. As a classic framework, 3D U-Net achieves good segmentation results. Based on the U-shaped network, 3D UX-Net introduces modules such as volumetric ConvNet with large kernel and improves model performance. Furthermore, Swin UNETR makes use of attention mechanism and achieves better results. As shown in Table \ref{exist_t}, our model has the optimal prediction accuracy, demonstrating that the methods proposed in this paper can effectively combine the advantages of CNNs and attention mechanism to improve the results of BMs segmentation task.

\begin{table}[htbp]
    \centering
    \caption{Performance comparison of various network architectures, including 3D U-Net, 3D UX-Net, Swin UNETR and our model.}
    \label{exist_t}
		\setlength{\tabcolsep}{1mm}{
    \begin{tabular}{c|cccc|cccc}
          \toprule
          \multirow{2}{*}{Model}& \multicolumn{4}{c|}{DSC~$\uparrow$} & \multicolumn{4}{c}{HD95~$\downarrow$} \\
          \cmidrule{2-9}
          & ET & TC & WT & Mean & ET & TC & WT & Mean \\
          \cmidrule{1-9}
          3D U-Net & 0.7170 & 0.7614 & 0.7565 & 0.7450 & 21.28 & 23.01 & 21.41 & 21.9 \\ 
          3D UX-Net & 0.7205 & 0.7616 & 0.7618 & 0.7480 & 19.97 & 21.65 & 16.94 & 19.52 \\ 
          Swin UNETR & 0.7263 & 0.7679 & 0.7753 & 0.7565 & 18.05 & 18.73 & 14.57 & 17.12 \\ 
          FANCL & \textbf{0.7563} & \textbf{0.8162} & \textbf{0.8027} & \textbf{0.7917} & \textbf{14.89} & \textbf{15.44} & \textbf{13.91} & \textbf{14.75} \\ 
          \bottomrule
    \end{tabular}}
\end{table}

\begin{figure}[htbp]
    \centering
    \includegraphics[width=0.8\linewidth]{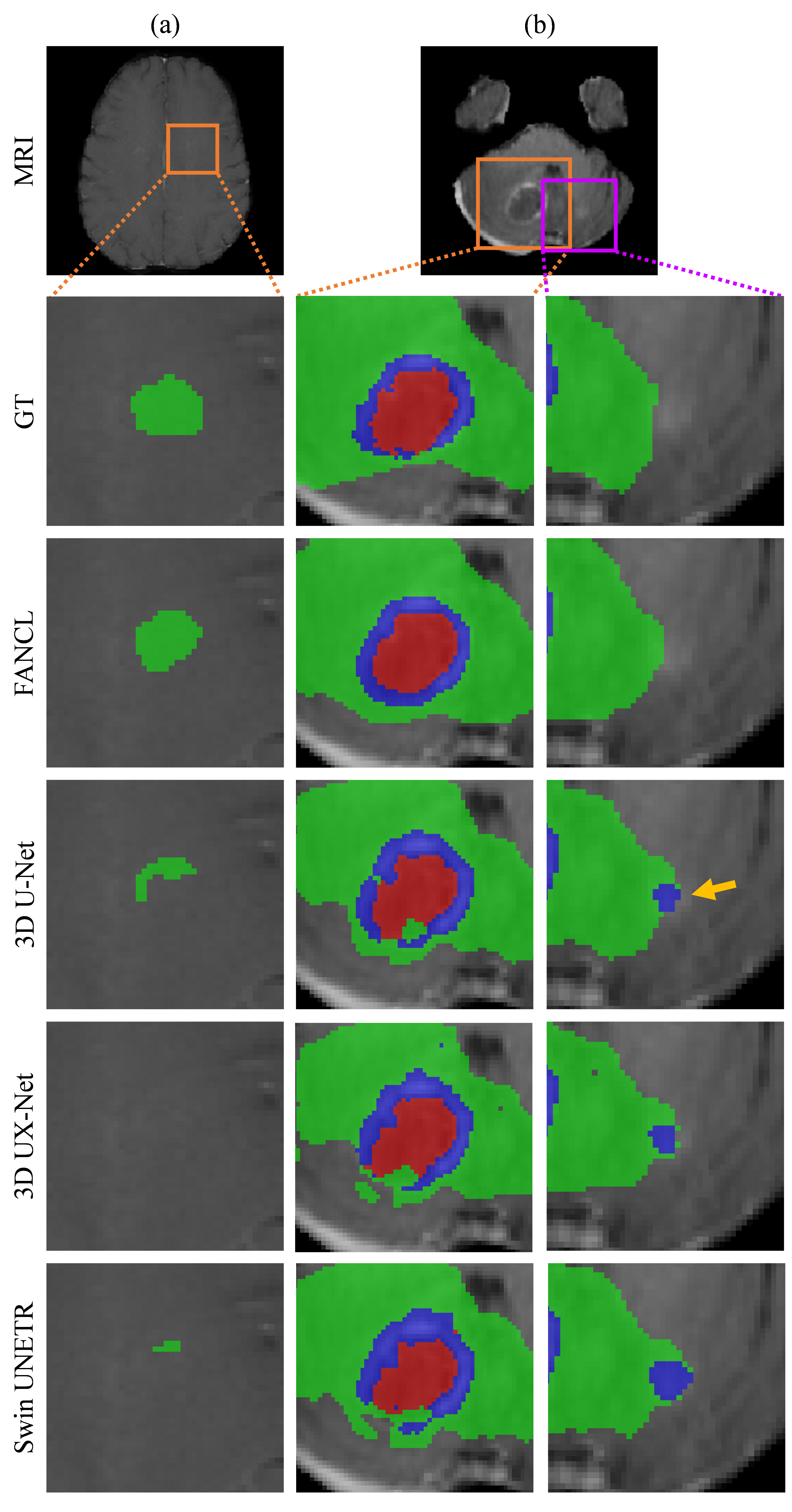}
    \caption{Comparative visualization of segmentation results across network architectures.}
    \label{exist_f}
\end{figure}

The relevant visualization results are shown in Fig. \ref{exist_f}. From (a), we can see that the prediction results of FANCL are much better than those of the existing approaches. In the tumor regions where the contrast with normal tissue is not significantly different, the outputs of the existing methods differ greatly from the ground truth, while FANCL reduces the gap. As shown in the top row of (b), compared with the existing methods, FANCL has better segmentation for the structure of BMs and the shape of each region. At the same time, FANCL also avoids erroneous prediction of the brighter part (see the yellow arrow in the bottom row of (b)), which is identified as the enhancing tumor in other methods.

\subsubsection{Ablation Experiments}
In this section, we conduct ablation studies on the feature-guided attention mechanism and CL strategy, and present the corresponding results.

\begin{table}[htbp]
    \caption{Ablation studies on the components of FANCL. C and F denote CL strategy and feature-guided attention, respectively.}
    \label{ablation}
    \centering
		\setlength{\tabcolsep}{1mm}{
    \begin{tabular}{c|cccc|cccc}
          \toprule
          \multirow{2}{*}{Model}& \multicolumn{4}{c|}{DSC~$\uparrow$} & \multicolumn{4}{c}{HD95~$\downarrow$} \\
          \cmidrule{2-9}
          & ET & TC & WT & Mean & ET & TC & WT & Mean \\
          \cmidrule{1-9}
          Baseline & 0.7449 & 0.8023 & 0.7754 & 0.7742 & 17.96 & 18.50 & 23.54 & 20.00 \\
          $\mathrm{Baseline+C}$ & 0.7540 & 0.8081 & 0.7870 & 0.7830 & 16.45 & 15.91 & 18.65 & 17.00 \\ 
          $\mathrm{Baseline+F}$ & 0.7530 & 0.8076 & 0.7958 & 0.7855 & 16.86 & 16.48 & 18.25 & 17.20 \\ 
          FANCL & \textbf{0.7563} & \textbf{0.8162} & \textbf{0.8027} & \textbf{0.7917} & \textbf{14.89} & \textbf{15.44} & \textbf{13.91} & \textbf{14.75} \\ 
          \bottomrule
    \end{tabular}}
\end{table}

$\textcircled{1}$ \textbf{Feature-guided Attention Mechanism:} We evaluate the performance of FANCL without attention mechanism (i.e. $\mathrm{baseline+C}$, the baseline trained with CL strategy). It can be seen in Table \ref{ablation} that compared with FANCL, the DSC values of each segmentation region show a significant decrease, while HD95 are increasing. This indicates that the feature-guided attention mechanism provides effective guidance for the segmentation of small tumors and improves the overall accuracy.

$\textcircled{2}$ \textbf{CL Strategy:} We train our model adopting the common training scheme to verify the effectiveness of CL strategy (i.e. $\mathrm{baseline+F}$, the baseline with feature-guided attention mechanism). As shown in Table \ref{ablation}, FANCL outperforms $\mathrm{baseline+F}$ in all three regions, especially in WT, where HD95 is reduced by 4.34, demonstrating that the progressive curriculum can effectively enhance the model's capacity to learn tumor structural features, thereby obtaining prediction results closer to the ground truth.

At the same time, we can also regard each ablation experiment as proof of the effectiveness of another module, because they both significantly improve the segmentation performance of baseline.

\section{Discussion}
This paper introduces a novel feature-guided attention mechanism integrated with a voxel-level CL strategy, achieving optimal performance in multi-class segmentation of multimodal BMs images. BMs are a particularly challenging category of tumor for automated segmentation. The significant variations in their presentation - occurring as either solitary or multiple lesions, with marked diversity in morphology and size - have posed substantial challenges for even state-of-the-art segmentation methods. Our study addresses these challenges through a carefully designed model that adopts a more appropriate structure and training strategy. The following comprehensive analysis encompasses model performance evaluation, negative case analysis, and limitations and future works.

\subsection{Model Performance Analysis}
\begin{figure}[htbp]
    \centering
    \includegraphics[width=0.3\linewidth]{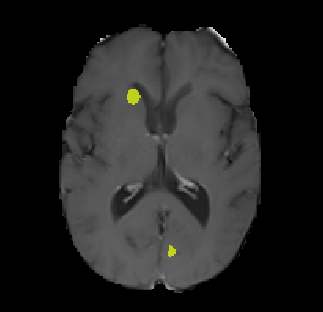}
    \caption{Example of the ground truth mask for the ET region, where the yellow part accounts for a relatively small proportion.}
    \label{ET_example}
\end{figure}

As can be seen from Section 5, FANCL demonstrates superior performance compared to all other models across both DSC and HD95 metrics. The ablation studies further validate the effectiveness of the proposed module. Specifically, a noteworthy observation from Table \ref{baseline_t} is the substantial performance improvement in the WT region. This enhanced performance can be attributed to two key factors: $\textcircled{1}$ Region and feature size. The WT region uniquely encompasses label 2 (peritumoral edema), which forms a significantly larger area compared to ET and TC regions. This increases the difficulty of FANCL guiding the segmentation of small tumors in ET and TC via large tumor features, because even the larger targets in ET and TC regions remain comparatively small and cannot provide sufficient information (as evidenced in Fig. \ref{ET_example}, where the ET region is highlighted in yellow). $\textcircled{2}$ Boundary contrast characteristics. The WT region exhibits more distinct boundary contrast compared to ET and TC regions, which is primarily due to the clear delineation between peritumoral edema and surrounding healthy brain tissue. The pronounced contrast facilitates more effective implementation of CL strategy, enabling more rational scheduling of curricula and providing enhanced progressive guidance to the model.

In addition, it should be noted that if there is a target foreground in the ground truth (even if it is very small) and the prediction only has the background, the maximum HD95 value of 373.1287 will be obtained. This situation frequently arises with models like nnU-Net, which tends to give conservative results in some complex samples, that is, a completely empty prediction output. In particular, since the model in WT task is trained on many images with relatively larger target regions, it will hesitate to give prediction when fed with some samples that only contain very small BMs, resulting in the maximum value of HD95. Therefore, this explains why the existing methods (Table \ref{exist_t}) have lower HD95 compared to baseline in WT task (Table \ref{baseline_t}), even though the DSC value of baseline is higher at this time. It is worth mentioning that despite the characteristic of the baseline architecture, FANCL still obtains the lowest HD95 value of 13.91 in WT region. This indicates that our proposed method does effectively improve the learning ability of the model.

\subsection{Analysis About Negative Cases}
In this study, two primary factors significantly influence FANCL's performance: image contrast characteristics, and complex morphology and shape patterns of metastatic lesions.

\begin{figure}[htb]
    \centering
    \includegraphics[width=0.8\linewidth]{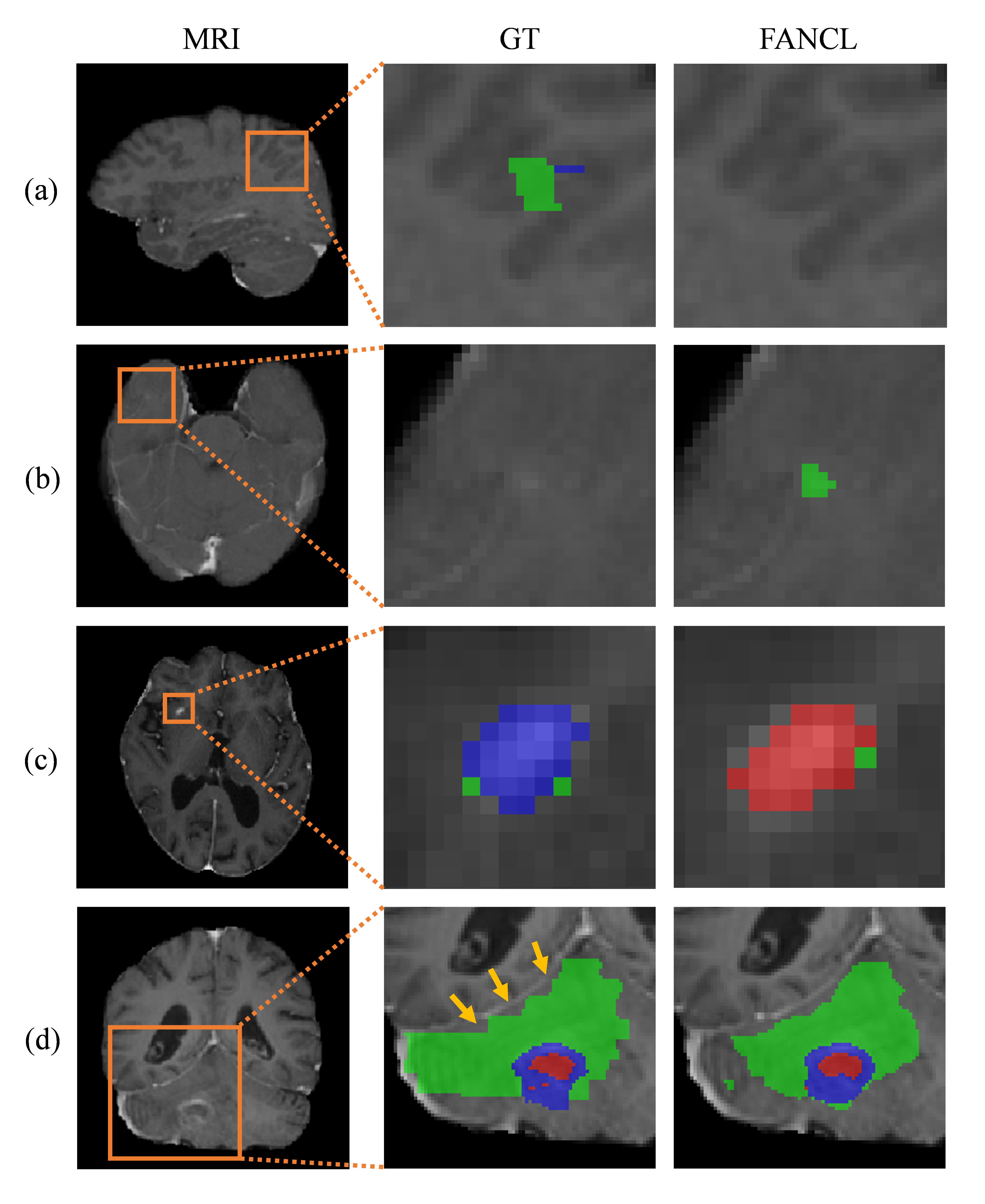}
    \caption{Examples of incorrect predictions.}
    \label{bad_examples}
\end{figure}

\subsubsection{Contrast-Related Challenges}
Our model exhibits specific performance limitations under certain contrast conditions. For example, in the low contrast scenarios as illustrated in Fig. \ref{bad_examples}(a), FANCL may fail to detect BMs when target regions show minimal contrast differentiation from surrounding tissues. This limitation particularly affects regions where tumor boundaries blend with adjacent tissue. Another example is false positive identifications. FANCL occasionally misidentifies locally hyperintense tissue regions as tumors. Fig. \ref{bad_examples}(b) demonstrates this limitation, where high-contrast areas are erroneously classified as peritumoral edema.

\subsubsection{Complex Morphology and Shape Patterns of Metastatic Lesions}
When segmenting some difficult samples, our model can provide preliminary predictions about the location and shape of BMs (as shown in Fig. \ref{bad_examples}(c)). However, due to the relatively small size of the tumor, FANCL cannot obtain sufficient contextual information from surrounding voxels, leading to misinterpretation of the enhancing tumor as a nonenhancing tumor. Furthermore, our model tends to yield predictions with continuously varying boundaries. Therefore, when faced with the tumor region as shown in Fig. \ref{bad_examples}(d), where the segmentation boundary of the ground truth is very neat and has many flat truncation (as indicated by yellow arrows), FANCL provides the boundary prediction with a slope.

\subsection{Limitations and Future Works}
Although the proposed method significantly improves the segmentation performance of BMs, certain challenging cases remain difficult to segment accurately, as shown in Fig. \ref{bad_examples}. To address these limitations, we plan to expand our training dataset and refine the attention mechanisms in future work. Furthermore, we will investigate the impact of different numbers of curriculum stages on model performance and analyze the effectiveness of FGA generated from different depths of the encoder. In addition, we will also explore the applicability of these methods to other models and tasks.

\section{Conclusion}
This paper presents FANCL, a novel model for brain metastases segmentation, that incorporates two key innovations, i.e., feature-guided attention mechanism and voxel-level curriculum learning strategy. The former leverages features from larger tumors to enhance small tumor segmentation, strengthens the model's capacity to capture global contextual information and optimizes feature utilization across tumors of varying sizes. The latter provides progressive guidance during model training, and enhances segmentation accuracy of tumor structure. Meanwhile, this paper implements an efficient curriculum-mining scheme to arrange curricula that conform to the model performance, which avoids model bias and maintains architectural simplicity. Finally, FANCL is evaluated on the BraTS-METS 2023 dataset, and the results show that the proposed method significantly improves the segmentation performance of baseline network and outperforms other methods, thereby demonstrating the effectiveness of FANCL.
\bibliographystyle{IEEEtran}
\bibliography{references}

\end{document}